# Induced robust topological order on an ordinary insulator hetero-structured with a strong topological insulator

**Bin Li,[1][†] Qiangsheng Lu,[2][†] Shuigang Xu,[3] Yipu Xia,[1] Wingkin Ho,[1] Ning Wang,[3] Chang Liu,[2][*] Maohai Xie[1][*]**

[1]*Department of Physics, The University of Hong Kong, Pokfulam Road, Hong Kong, China.*

[2]*Department of Physics, Southern University of Science and Technology, Shenzhen, Guangdong, 518055, China.*

[3]*Department of Physics, The Hong Kong University of Science and Technology Clear Water Bay, Kowloon, Hong Kong, China.*

**Topological states of matter originate from distinct topological electronic structures of materials.[1–6] As for strong topological insulators (STIs), the topological surface (interface) is a direct consequence of electronic structure transition between materials categorized to different topological genus.[7,8] Therefore, it is fundamentally interesting if such topological character can be manipulated. Besides tuning the crystal field and the strength of spin-orbit coupling (e.g., by external strain,[9–12] or chemical doping[13]), there is currently rare report on topological state induced in ordinary insulators (OIs) by the heterostructure of OI/STI.[14] Here we report the observation of a Dirac cone topological surface state (TSS) induced on the $Sb_2Se_3$ layer up to 15 nm thick in the OI/STI heterostructure, in sharp contrast with the OI/OI heterostructure where no sign of TSS can be observed. This is evident for an induced topological state in an OI by heterostructure.**

[†]These authors contributed equally to this work. *Email: liuc@sustc.edu.cn and mhxie@hku.hk

Both $Sb_2Se_3$ and $Bi_2Se_3$ belong to the family of V(Bi, Sb)-VI(Se, Te) compounds.[15] Among the four binary compounds in this family, three are predicted and confirmed to be STIs.[16–18] The exception for $Sb_2Se_3$ to be an OI is due to the weaker spin orbit coupling. There are currently three available regimes in which the topological phase transition of $Sb_2Se_3$ may occur. First, it was suggested that external strain can transform the rhombohedral phase of $Sb_2Se_3$ from an OI into an STI,[10,11] evident by a recent Raman study,[19] but successive experiment casted doubt on whether the rhombohedral phase is realized there.[20] Second, a recent report has claimed that the stand-alone $Sb_2Se_3$ is readily a STI.[21] Third, a first-principles study[14] suggested that STI can be induced at the interface of $Sb_2Se_3/Bi_2Se_3$ for both materials, i.e., $Sb_2Se_3$ transforms from an OI to a STI, and the opposite holds for $Bi_2Se_3$. In this letter, we will comment on these theoretical proposals and give our assumption.

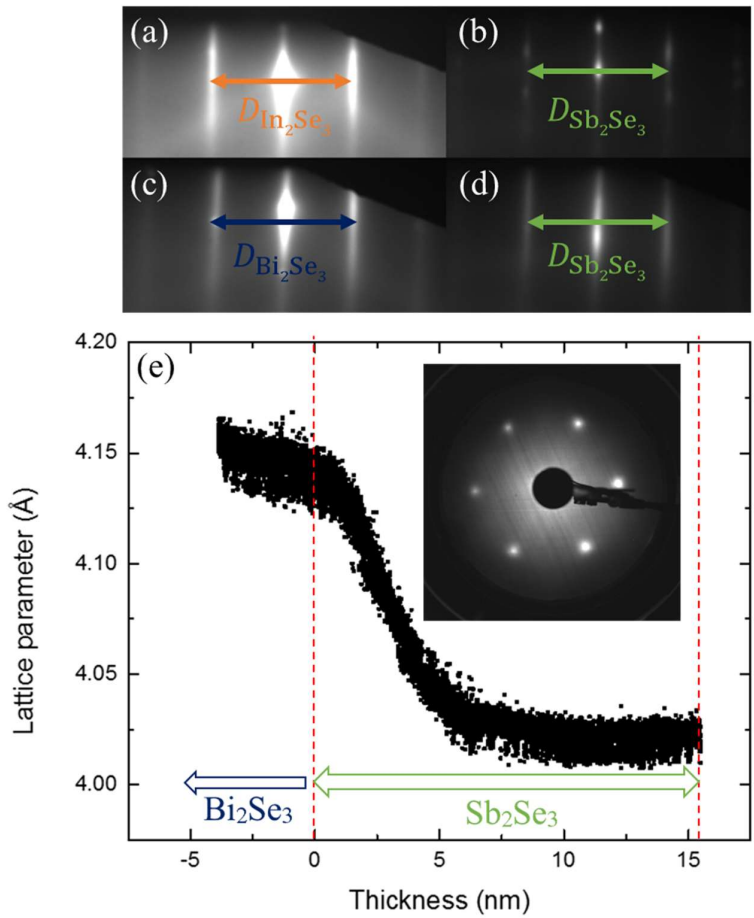

Figure 1: Growth characterizations for the OI/STI and OI/OI heterostructures by RHEED and LEED. (a) and (c) are RHEED patterns of $In_2Se_3$ and $Bi_2Se_3$ respectively. (b) and (d) are patterns of $Sb_2Se_3$. Orange, blue and green arrows indicate the spacings measured. (e) Lattice parameter variation during $Sb_2Se_3$ growth. Two red dashed lines bracket the period of $Sb_2Se_3$ growth as marked by the green arrow. Before the first red line it is the lattice parameter of $Bi_2Se_3$ under thermal stabilization. LEED pattern (the inset) of the grown $Sb_2Se_3$ shows the symmetry being consistent with the rhombohedral lattice.

In order to obtain a STI phase in $Sb_2Se_3$ either by strain or hetero-structuring, a prerequisite is to grow this material in the *rhombohedral* phase (Supplementary file) instead of the stable orthorhombic phase commonly used in experiments.[22] To start the growth process, thick

In$_2$Se$_3$ or Bi$_2$Se$_3$ films were firstly grown on InP(111), showing good crystal quality of both according to the electron diffraction measurements. For example, Figure 1a and 1c are the RHEED patterns taken from the grown In$_2$Se$_3$ and Bi$_2$Se$_3$ surfaces respectively. The spacings between $(01)$ and $(0\bar{1})$ diffraction streaks (i.e., $D_{In_2Se_3}$ and $D_{Bi_2Se_3}$ as marked by the orange/blue arrows in Figure 1a and 1c) are found to correspond well with the in-plane lattice constants of strain-free In$_2$Se$_3$ and Bi$_2$Se$_3$ crystals. On top of such In$_2$Se$_3$ and Bi$_2$Se$_3$ layers, Sb$_2$Se$_3$ up to 15 QLs thick are deposited and the corresponding RHEED patterns are shown in Figures 1b and 1d. Again, the inter-diffraction streak spacings were measured, which were translated into an in-plane lattice constant of 4.02 Å, agreeing well with the theoretically optimized lattice constants of rhombohedral Sb$_2$Se$_3$.[10,11,15,21] In fact, the lattice parameter evolution has been followed by RHEED during Sb$_2$Se$_3$ deposition and Figure 1e represents an example for the case of Sb$_2$Se$_3$ growth on Bi$_2$Se$_3$. As it is evident, the lattice changes from that of the Bi$_2$Se$_3$ 'substrate' to that of Sb$_2$Se$_3$ during the first 5 nm deposition period. For a film as thick as 15 nm, the lattice constant stabilizes at 4.02 Å and is interpreted as strain-free. This is crucial since it excludes the possibility that the observed topological state was introduced by substrate strain. The LEED pattern of the grown Sb$_2$Se$_3$ (the inset of Figure 1e) clearly shows the 6-fold hexagonal symmetry, evidencing the rhombohedral structure of Sb$_2$Se$_3$.

In pursuit of more detailed structural information of the sample, XRD measurements were performed. The diffraction patterns of a pure Bi$_2$Se$_3$ film and samples with Sb$_2$Se$_3$ grown on top were compared. A new set of diffraction peaks emerged close to the corresponding ones of rhombohedral Sb$_2$Se$_3$ lattice, with the same ratio of lattice-plane distances (d$_{Sb2Se3}$ : d$_{Bi2Se3}$ ≈ 1.1), thus the rhombohedral Sb$_2$Se$_3$(00$x$) diffraction peaks are clearly identified. Figure 2a present an example of the resulted $\theta - 2\theta$ scans (see Supplementary Figure S2 for a scan over a larger angle range) of a pure Bi$_2$Se$_3$ film and a Sb$_2$Se$_3$/Bi$_2$Se$_3$ heterostructured sample, respectively, both grown on InP. The new peak aroused from (0, 0, 18) lattice planes of the rhombohedral Sb$_2$Se$_3$ in Fig. 2a is evident. The lattice constant along the *c*-axis of epitaxial Sb$_2$Se$_3$ is ~31.5 Å according to the XRD measurement, in good agreement with the theoretical value of rhombohedral Sb$_2$Se$_3$.[10,11,15,21] If the Sb$_2$Se$_3$ film were of the orthorhombic phase, the lattice constant would be $a = b \approx 12$ Å and $c = 4$ Å, far from the experimental values. In the XRD data, hexagonal Bi$_2$Se$_3$ as a reference with lattice constant of $c = 28.6$ Å is also well recognized and in agreement with previous reports.[15]

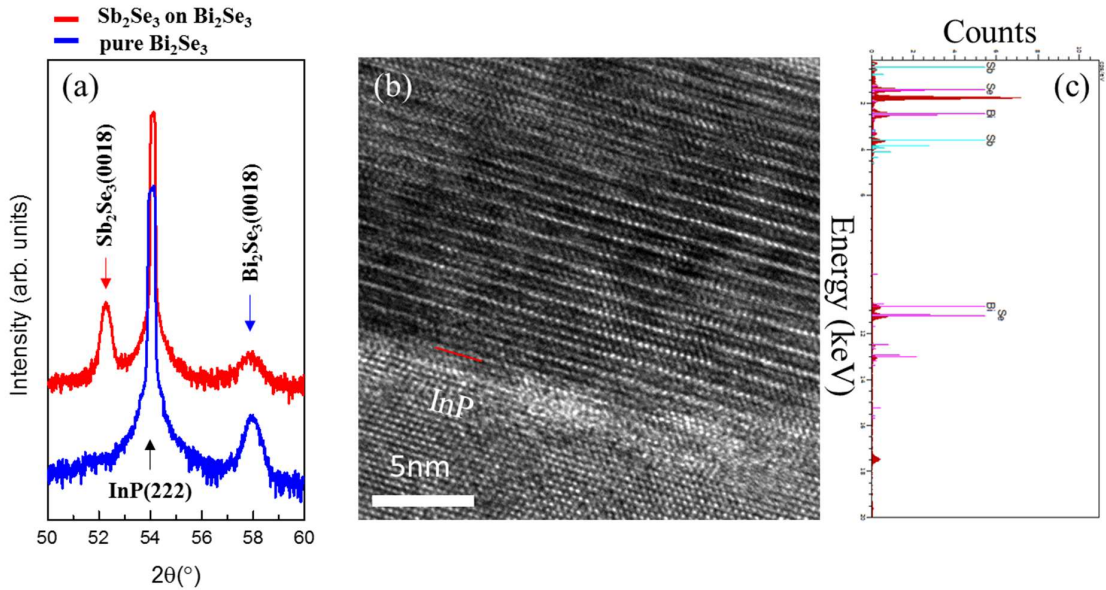

Figure 2: XRD and TEM characterizations. (a) XRD θ-2θ scan curves of a pure Bi$_2$Se$_3$ film (blue curve) and a Sb$_2$Se$_3$-on-Bi$_2$Se$_3$ heterostructured sample (red curve), both grown on InP(111) substrate. (b) Cross sectional HRTEM image and (c) the EDS result of the Sb$_2$Se$_3$/Bi$_2$Se$_3$ heterostructured sample.

Cross sectional HRTEM image of the Sb$_2$Se$_3$/Bi$_2$Se$_3$ sample is shown in Figure 2b. From the image, one only observes a single phase despite the sequential growth of Bi$_2$Se$_3$ and Sb$_2$Se$_3$. Thus, it supports the RHEED/LEED observation that the Sb$_2$Se$_3$ lattice structure is the same as that of the rhombohedral $Bi_2Se_3$. To show that $Sb_2Se_3$ does have grown, energy dispersive X-ray spectroscopy (EDS) has been taken from the epifilm, which indeed reveals Sb, Bi and Se elements (see Figure 2c).

The electronic structure of rhombohedral $Sb_2Se_3$ has not been observed by experiment so far. In this work, topological electronic structure of $Sb_2Se_3$ grown on $Bi_2Se_3$ is captured by ARPES even for samples with a thick $Sb_2Se_3$ epifilm (~15 nm). For comparison, the same film grown on $In_2Se_3$, an OI, reveals no topological state. This experiment thus provides the first experimental evidence of the non-trivial topological state in $Sb_2Se_3$.

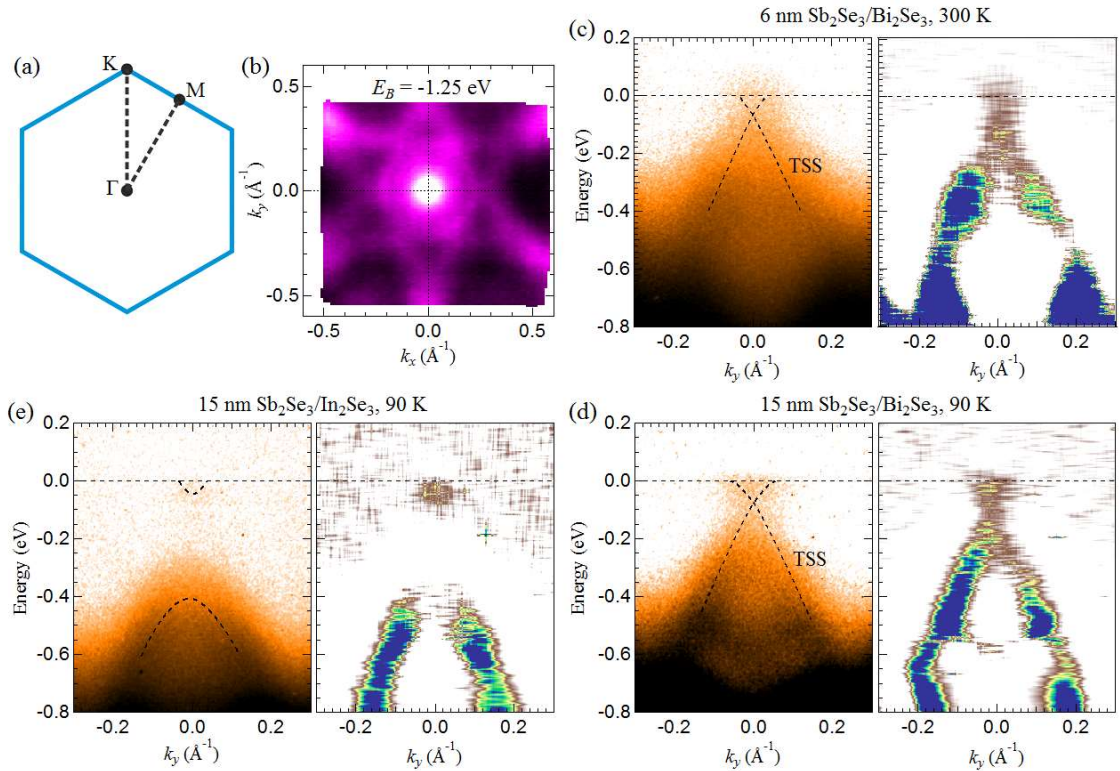

Figure 3 ARPES results from the heterostructures. (a) Surface Brillouin zone and high symmetry points. (b) Constant energy mapping at $E_B$ = -1.25 eV. (c) ARPES results of 6 nm $Sb_2Se_3$-on-$Bi_2Se_3$ (OI/STI) heterostructure measured at 300 K. Right panel shows the band dispersion under second derivation data processing [same in (d) and (e)]. Both panel shows a clear Dirac cone TSS (dashed line). (d) ARPES data of 15 nm $Sb_2Se_3$-on-$Bi_2Se_3$ (OI/STI) measured at 90 K. The Dirac cone TSS still exists despite the absence of in-plane strain. (e)

ARPES data on 15 nm $Sb_2Se_3$-on-$In_2Se_3$ (OI/OI) at 90 K as a control measurement. No Dirac cone is seen on this OI/OI structure.

The ARPES data is shown in Figure 3. In our measurements, two sets of samples with different $Sb_2Se_3$ layer thicknesses (6 nm and 15 nm) are compared. First, the constant energy mapping under -1.25 eV binding energy also showed the 6-fold symmetry for the bulk band (Figure 3b), agreeing with the LEED data and offers another evidence that the $Sb_2Se_3$ used in the experiment is rhombohedral. The direction of the horizontal $k$ is along K-Γ-K in the first Brillouin zone as illustrated in Figure 3a. Figure 3c is the result for the 6 nm-$Sb_2Se_3$ grown on $Bi_2Se_3$. A cross-like state connecting the conduction band and the valence band is resolved, which is a prominent feature of the Dirac cone TSS. It should be noted that 6 nm is already thick enough to exclude the smearing effect of topological state extending from $Bi_2Se_3$ to $Sb_2Se_3$. To further verify that the topological state is from $Sb_2Se_3$, thicker samples are prepared and measured. Figure 3d exhibits the electronic structure of a 15 nm-thick $Sb_2Se_3$ grown on $Bi_2Se_3$. Again, the same 'cross-like' state is observed. This TSS can be seen more clearly by the second derivation processing as shown in the insets in Figure 3c-3e. Importantly, no such 'cross-like' TSS can be discerned for a similar 15 nm-$Sb_2Se_3$ film but grown on $In_2Se_3$ (Figure 3e and there is almost no misfit between two lattices). Therefore, the interpretation of intrinsic topological state is also less relevant because the distinction of electronic structures between $Sb_2Se_3$ epifilms grown on $Bi_2Se_3$ and $In_2Se_3$ is outstanding. Based on all the evidences above, it can be concluded that the 'cross-like' state is the non-trivial topological surface state and what have been observed in Figure 3e contains only topologically trivial states. The Dirac point is located at ~0.1 eV below the Fermi level. For the trivial state, a gap of more than 0.25 eV is determined by the ARPES data.

There is now only one regime left that such topological state is likely induced by either the proximity effect or some more involved mechanism. The proximity effect by itself holds great promises of quite some exotic phenomena such as the topological superconducting phase (Majorana fermion),[23] magnetic monopoles,[24] quantum anomalous Hall effect,[25] etc., making this procedure by itself of great scientific interest. Currently, the proximity effect induced topological phase has been mainly studied for the interface of bismuth bilayer and a STI,[26] but it has also been reported that bismuth itself may be a STI;[27,28] Bi bilayer is also too thin to experimentally discriminate signals from the STI substrate. Our system, on the other hand, is valid to check this theorem. According to the analysis in Ref. [14], it is because the reordering of Se energy level at the interface, which leads to the sign change of parity values for both $Sb_2Se_3$ and $Bi_2Se_3$. Such effect, if real, would be a completely new approach to manipulate the topological state. We tend to assign this regime to our observation. Nevertheless, it still remains unclear whether the proximity effect can affect the $Sb_2Se_3$ with a thickness of ~15 nm, as well as its relationship with the predicted 'bulk-like' topological state, thus this is inviting further theoretical work.

**Methods**

Film growth and some surface characterizations were carried out in a multi-chamber ultrahigh vacuum (UHV) facility with the background pressure of ~$10^{-10}$ torr. A customized Omicron MBE reactor was employed to grow the samples, where elemental sources of Sb, Bi, In and Se were installed in Knudsen cells. The flux ratio among Sb, Bi, In and Se was set at 1.25:1:1:10. The growth temperatures for $Bi_2Se_3$, $In_2Se_3$, and $Sb_2Se_3$ were 200 °C, 500 °C, and 150 °C, respectively, and the growth rates were estimated by post-growth measurements of film thickness and deposition time. The growing surfaces were monitored in real-time by reflection high energy electron diffraction (RHEED) operated at 10 keV. A high-speed CCD camera was utilized to capture the RHEED pattern and to extract the lattice constant information. After the growth, the samples were transferred to an adjacent low energy electron diffraction (LEED) chamber for surface lattice analysis. The crystal structure of $Sb_2Se_3$ was also characterized by X-ray diffraction (XRD) symmetrical θ-2θ scans. The $Sb_2Se_3$/$Bi_2Se_3$ interface was characterized by high-resolution transmitted electron microscopy (HRTEM) using a JEOL 2010F TEM microscope operated at 200 keV. EDS measurements were performed in the TEM with the same electron energy and the electron beam were set to sample on the whole region of the deposit.

Angle-resolved photoemission spectroscopy (ARPES) measurements were performed using a laboratory-based ARPES system consisted of a SPECS PHOIBOS 150 electron analyzer and a UVLS-600 UV lamp under the pressure of $3 \times 10^{-10}$ mbar. The thin film samples were capped with Se protection layers before they were decapped in the ARPES chamber with a pressure below $2 \times 10^{-9}$ mbar. Decapping temperature was 160~200 °C measured by an infrared temperature gun. *In situ* Auger electron spectroscopy measurements were done after decapping to ascertain the chemical elements on the surface. The incident photon energy was 21.218 eV (He I), the spot diameter was about 500 micrometers, and the measuring temperatures were ~90 K and room temperature, with an energy resolution of 40-50 meV at 90 K. Samples were found to be stable during a typical measurement period of ~20 hours.

Acknowledgements

The work described in this paper was supported in parts from a grant of the SRFDP and RGC ERG Joint Research Scheme of Hong Kong Research Grant Council (RGC) and the Ministry of Education of China (No. M-HKU709/12) and from a Collaborative Research Fund (HKU9/CRF/13G) sponsored by the RGC of Hong Kong Special Administrative Region, China. Work at SUSTech was supported by Grant No. 11504159 of NSFC, Grant No. 2016A030313650 of NSFC Guangdong, and Project No. JCY20150630145302240 of the Shenzhen Sci. & Tech. Innovations Committee.

Author contributions

This study was initiated by B. L. and M. H. X. and the experiment was designed by B. L., Q. S. L., C. L., and M. H. X.; B. L., Y. P. X. and M. H. X. contributed to crystal growth and structural characterizations by the RHEED, LEED, and XRD. Q. S. L. and C. L. contributed

# Supplementary Materials

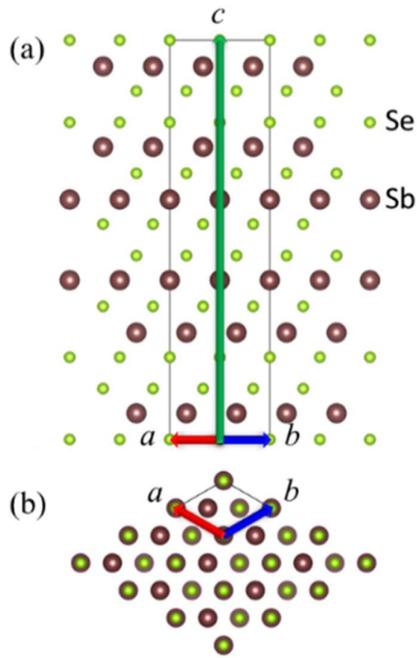

Figure S1: The crystal structure. (a) Side (a) and (b) top view of rhombohedral $Sb_2Se_3$ crystal. The black frames exhibit the unit cell with the basis vectors of *a*, *b* and *c*. The atoms of Sb and Se have been labeled in (a).

The crystal structure of rhombohedral phase of $Sb_2Se_3$ is illustrated in Figure S1 both from the side (a) and the top views. The black frames represent the unit cell with the basis vectors of *a*, *b* and *c*. The atoms of Sb and Se have also been labeled in (a). As elaborated in the main text, we achieved this by MBE growth on both $Bi_2Se_3$ and $In_2Se_3$ films deposited on InP(111). The latter two crystals are both of the rhombohedral structure in equilibrium, whereas rhombohedral $Sb_2Se_3$ is metastable.

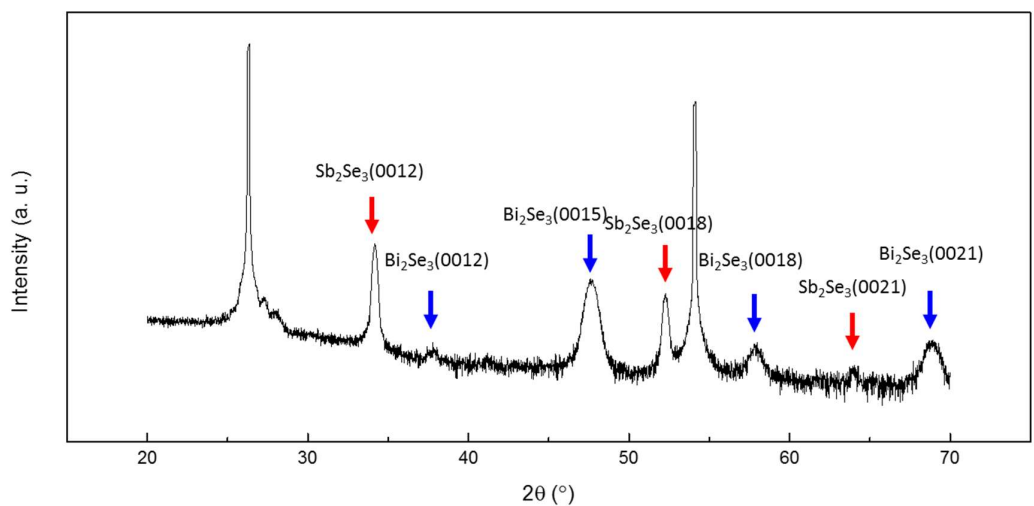

Figure S2: XRD θ-2θ scan of a $Sb_2Se_3/Bi_2Se_3$ heterostructured sample. The diffraction peaks from rhombohedral $Sb_2Se_3$ and $Bi_2Se_3$ are indicated by red and blue arrows, respectively. The unlabeled strong peaks are for InP substrate.

XRD θ-2θ scan over a large angle range of a $Sb_2Se_3/Bi_2Se_3$ heterostructured sample is presented in Figure S2, which reveals the various diffraction peaks (indicated by arrows) of rhombohedral $Sb_2Se_3$ and $Bi_2Se_3$. These peaks can be consistently assigned to the rhombohedral phase of $Sb_2Se_3$, $Bi_2Se_3$, and InP substrate.

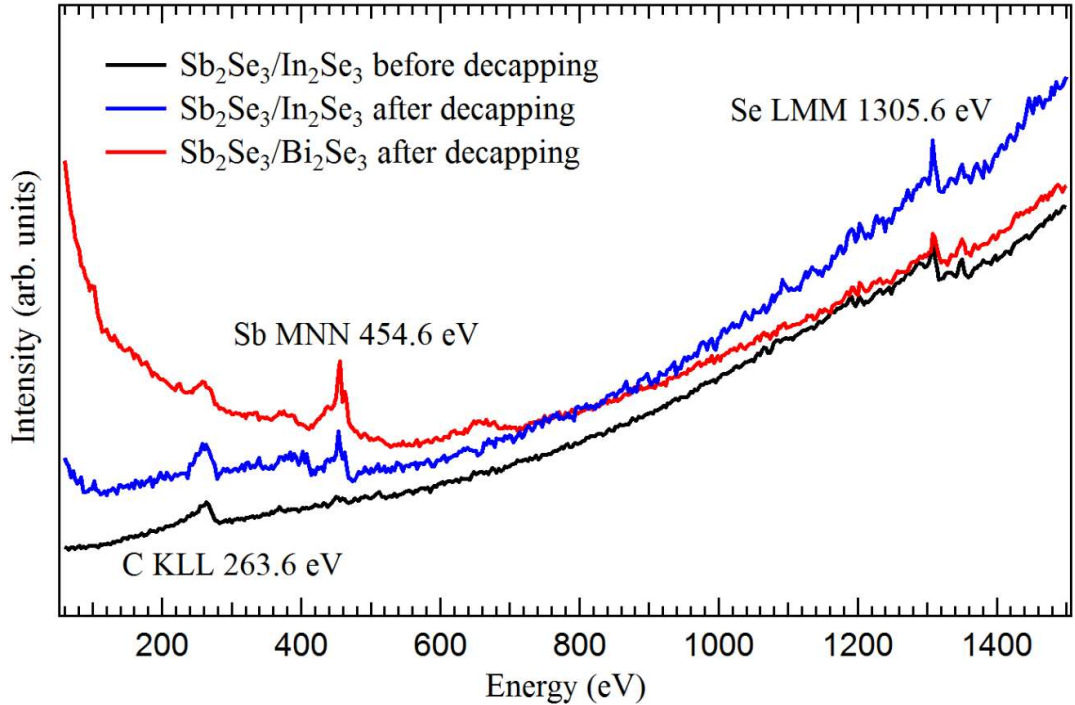

Figure S3: Auger electron spectra for heterostructured $Sb_2Se_3/Bi_2Se_3$ and $Sb_2Se_3/In_2Se_3$ samples before and after the *in-situ* decapping of the Se protection layer. Before decapping, the film surface contained only Se and the residue C component (black curve), while the Sb MNN peak is clearly visible for both $Sb_2Se_3/Bi_2Se_3$ (red curve) and $Sb_2Se_3/In_2Se_3$ (blue curve) samples after decapping, suggesting a clean $Sb_2Se_3$ surface.

As described in the main text, we used a 'decapping' procedure for surface preparation inside the ARPES vacuum chamber before the ARPES measurements. Before these films were loaded to the ARPES chamber, they were capped with Se protection layers to avoid oxidation and degradation of the $Sb_2Se_3$ surfaces. After loading, we heated up the heterostructures to about 160~200 °C such that only the Se protection layer but not the $Sb_2Se_3$ layer was vaporized. Figure S3 shows the Auger electron spectra for the heterostructures before and after decapping. Before decapping, the film surface contained only Se and the residual C component (black curve), while the Sb MNN peak is clearly visible for both the OI/STI (red curve) and the OI/OI heterostructure (blue curve) after decapping. Therefore, a clean Sb$_2$Se$_3$ layer is recovered by the decapping procedure.